\documentclass[11pt, reqno]{amsart}

\usepackage{amsmath,amssymb, mathabx}
\usepackage{graphicx}
\usepackage[pagebackref, colorlinks = true, linkcolor = blue, urlcolor  = blue, citecolor = red]{hyperref}
\usepackage{xcolor}

\usepackage[margin=1in]{geometry}

\usepackage{subcaption} 
\usepackage{colortbl}
\usepackage{multirow}

\usepackage{layout}
\usepackage{enumitem}
\usepackage{breqn}

\newcommand{\eq}[1]{\begin{align}\begin{aligned}#1\end{aligned}\end{align}}

\newtheorem{theorem}{Theorem}[section]
\newtheorem*{theorem*}{Theorem}

\newtheorem{corollary}[theorem]{Corollary}

\newtheorem{remark}[theorem]{Remark}

\DeclareMathOperator*{\E}{\mathbb E} 
\DeclareMathOperator{\trace}{Tr} 
\DeclareMathOperator{\owg}{oWg} 
\DeclareMathOperator{\uwg}{uWg} 

\graphicspath{{figures/}}
\numberwithin{equation}{section}

\date{}

\title{Symbolically integrating tensor networks over various random tensors by the second version of Python RTNI}

\author{Motohisa Fukuda}
\address{MF: Yamagata University, 1-4-12 Kojirakawa, Yamagata, 990-8560 Japan}
\email{fukuda@sci.kj.yamagata-u.ac.jp}

\begin{document}
\maketitle
\markright{\MakeUppercase{PyRTNI2}}

\begin{abstract}
We are upgrading the \texttt{Python}-version of \texttt{RTNI}, which symbolically integrates tensor networks over the Haar-distributed unitary matrices \cite{RTNI}. Now, \texttt{PyRTNI2} can treat the Haar-distributed orthogonal matrices and the real and complex normal Gaussian tensors as well. Moreover, it can export tensor networks in the format of \texttt{TensorNetwork} \cite{roberts2019tensornetwork} so that one can make further calculations with concrete tensors, even for low dimensions, where the Weingarten functions differ from the ones for high dimensions. 
The tutorial notebooks are found at \texttt{GitHub}: \url{https://github.com/MotohisaFukuda/PyRTNI2}. 
In this paper, we explain maths behind the program and show what kind of tensor network calculations can be made with it. For the former, we interpret the element-wise moment calculus of the above random matrices and tensors in terms of tensor network diagrams, and argue that the view is natural, relating delta functions in the calculus to edges in tensor network diagrams. 
\end{abstract}

\section{Introduction}
Calculating moments of Haar-distributed unitary matrices plays an important role in quantum physics because of the unitary evolution. Especially, in quantum information theory, assessing typical behaviours of systems led to fruitful insights \cite{hayden2006aspects, collins2016random}. 
We refer to our precious paper \cite{RTNI} for more relations of random matrices and random tensor networks to quantum physics. 
The integral of a monomial over the Haar-distributed unitary matrices:
\begin{align}\label{eq:monomial}
\E_{U \sim \mathcal U(d)} \left[u_{i_1, j_1}\cdots u_{i_k, j_k} \cdot
\overline{u_{i^\prime_1, j^\prime_1}\cdots u_{i^\prime_k, j^\prime_k}} \right]
\end{align}
was investigated in the asymptotic regime as $d \to \infty$ in \cite{weingarten1978asymptotic}, 
and for finite dimensions in \cite{CollinsSniady2006}, which gave the exact combinatorial formula called Weingarten calculus involving the Weingarten functions which are algebraically defined.
This calculus was implemented symbolically as \texttt{RTNI} \cite{RTNI} with \texttt{Mathematica} and \texttt{Python}, for tensor networks with some of nodes being Haar-distributed unitary matrices of symbolic dimensions. So far, \texttt{RTNI} has been used for quantum neural networks and quantum circuits \cite{cerezo2021cost, pesah2021absence, sharma2022trainability, liu2023analytic, wang2022symmetric, liu2022mitigating}, more broadly quantum information and quantum physics \cite{fukuda2019typical, liu2020scrambling, kukulski2021generating, riddell2021relaxation, iosue2023page, haag2023typical, cipolloni2023non} and mathematics \cite{tarrago2020spectral}.

Now, we make further steps to develop \texttt{Python}-based \texttt{PyRTNI2}. First, it accommodates not only the unitary matrices, but also the orthogonal matrices, the random complex and real normal Gaussian tensors, where the unitary and orthogonal matrices are equipped with normalized Haar measures, which are unique invariant measures apart from the normalization. It also provides the Weingarten functions for low dimensional cases. Element-wise moments of the unitary, orthogonal and symplectic matrices can be calculated by the \texttt{Maple}-based \texttt{IntHaar} \cite{ginory2021weingarten} and element-wise moments of the unitary matrices even for low dimensions by the \texttt{Mathematica}-based \texttt{IntU} \cite{puchala2017symbolic}. Nevertheless,  \texttt{PyRTNI2} deals with matrices and tensors symbolically, which gives a great advantage while integrating tensor networks. 
When integration is performed over the unitary group $\mathcal U(d)$ or the orthogonal group $\mathcal O(d)$, one needs to take account of Weingarten functions \cite{CollinsSniady2006, CollinsMatsumoto2009}, which are defined by the dimension $d$ of the random matrix and the size $k$ of symmetric group $S_k$ which is relevant to the number of copies of the random matrix; see Section \ref{section:weingarten} for the detail. However, as in Remark \ref{remark:low-dim}, individual cases for  $d<k$ require different formulae of Weingarten functions. \texttt{PyRTNI2} comes with Weingarten numbers for such pairs of $d$ and $k$, and can evaluate integrals for concrete dimensions. 
Next, the tensor networks with \texttt{PyRTNI2} can be exported in the format of \texttt{TensorNetwork} \cite{roberts2019tensornetwork} so that one can continue to make further tensor network calculations. 

Among the \texttt{Python} standard library modules, \texttt{os}, \texttt{itertools}, \texttt{json}, \texttt{copy} and \texttt{math} were used. 
For the symbolic calculations only one third-party module \texttt{SymPy} \cite{sympy} was needed to manipulate combinatorics and symbolic polynomials. On the other hand, one would need \texttt{TensorNetwork} \cite{roberts2019tensornetwork} and \texttt{NumPy} \cite{numpy_harris2020array}
to deal with concrete tensor networks and \texttt{Graphviz} \cite{ellson2002graphviz} to get their graphical representations via \texttt{TensorNetwork}. 

Now, let us explain our mathematical notations.
Elements of tensors are usually specified by sub-scripted indices as in \eqref{eq:monomial}, but from here on we use functional representation, for example, the $i$-th element of a vector $V$ and the $(i, j)$-element of a matrix $M$ are written as $v(i)$ and $m(i,j)$, respectively. This notation is also extended naturally to tensors with more axes. 
The Haar-distributed $n \times n$ orthogonal and unitary matrices are respectively denoted by $\mathcal O(n)$ and $\mathcal U(n)$. Also, the Gaussian distributions of $n_1 \times \cdots \times n_N$ real and complex tensors are respectively denoted by 
$\mathcal G_{\mathbb R} (n_1, \ldots, n_N)$ and $\mathcal G_{\mathbb C} (n_1, \ldots, n_N)$. The operations of transpose, complex conjugate and adjoint are expressed as 
$\cdot^T$, $\bar \cdot$ and $\cdot^*$, respectively.
The canonical basis in $\mathbb R^d$ are denoted as $\{|i \rangle \}_{i=1}^d$. When it is a multi-index, i.e. $i = (i_1, \ldots, i_{L})$, we understand that 
\begin{align*}
|i\rangle = |i_1\rangle \otimes \cdots \otimes |i_L\rangle \ .
\end{align*}
Moreover, to drop a vector, 
\begin{align*}
| i\{\ell\} \rangle = |i_1\rangle \otimes \cdots \otimes |i_{\ell-1}\rangle \otimes |i_{\ell+1}\rangle \otimes \cdots \otimes |i_L\rangle \ .
\end{align*}

As for combinatorics, $S_k$ stands for the permutation group of $k$ elements, and $P_{2k}$ pairings of $2k$ elements. An element $p \in P_{2k}$ can be written uniquely as
\eq{\label{eq:part-perm}
\{\{p(1), p(2)\}, \ldots ,\{p(2k-1), p(2k)\}\} \ ,
}
where $p(2\ell-1) < p(2 \ell)$ for $\ell \in [k]$ and $p(1) < \cdots <p(2k-1)$. 
Also, in this study each $p$ in \eqref{eq:part-perm}  will be regarded as an element of $S_{2k}$ in the following way:
\eq{
(p(1), p(2)), \cdots (p(2k-1), p(2k)) \ ,
}
otherwise stated. 
Complex and real Weingarten functions are denoted by $\uwg(\cdot, \cdot)$ and $\owg(\cdot, \cdot)$, whose definitions are found in Section \ref{section:weingarten}.

Tutorial notebooks are provided at \texttt{GitHub}, with which one can see how to use it through examples, while the current paper explains mathematics behind the calculus and possible applications of \texttt{PyRTNI2}.
Indeed, the element-wise moment formulae are presented in Section \ref{section:calculus},
the formulae are discussed in the viewpoints of tensor networks in Section \ref{section:tensornetworks} and further applications of \texttt{PyRTNI2} are discussed in Section \ref{section:further}. Moreover, Appendix \ref{section:weingarten} provide more details on Weingarten functions.

\section{Formulae for element-wise moments of random tensors}\label{section:calculus} 
This section explains how to calculate element-wise moments of random matrices and tensors of our interest:
the Haar-distributed unitary and orthogonal matrices, and the complex and real normal Gaussian vectors. 
Indeed, the Weingarten calculus is used for the first two \cite{CollinsSniady2006, CollinsMatsumoto2009} and the Wick formulae for the last two \cite{isserlis1918formula, wick1950evaluation, mingo-speicher-book}.
In addition, notice that realigning tensors, for example in the lexicographic order, yields vectors. As such, real and complex normal Gaussian tensors can be identified as vectors, while it is not the case with orthogonal and unitary matrices, where distinguishing the input and the output spaces is essential. In particular, the following claims on real and complex normal Gaussian vectors can be naturally extended to tensors. 
The readers can refer to Appendix \ref{section:weingarten} for more about the Weingarten functions: $\uwg(\cdot, \cdot)$ and $\owg(\cdot, \cdot)$.

\begin{theorem}\label{theorem:calculus}
Integrating monomials of elements of the unitary and orthogonal matrices, and the complex and real normal Gaussian vectors can be individually stated as follows. 
\begin{enumerate}
\item 
For a Haar-distributed $d \times d$ unitary matrix $M$, and $i_\ell, i^\prime_\ell, j_\ell, j^\prime_\ell \in [d]$ with $\ell \in [k]$,
\eq{
&\E_{M \sim \mathcal U(d)} \left[m(i_1, j_1)\cdots m(i_k, j_k) \cdot
\overline{m(i^\prime_1, j^\prime_1)\cdots m(i^\prime_k, j^\prime_k)} \right]\\
&=\sum_{\alpha, \beta \in S_k} 
\delta(i_1, i^\prime_{\alpha (1)}) \cdots \delta(i_k, i^\prime_{\alpha(k)}) 
\delta(j_1, j^\prime_{\beta(1)}) \cdots \delta(j_k, j^\prime_{\beta(k)}) 
\, \uwg(d, (\alpha, \beta)) \ .
}
When the numbers of $m(\cdot, \cdot)$'s and $\overline{m(\cdot, \cdot)}$'s are different, the integral is trivially zero. 

\item 
For a Haar-distributed $d \times d$ orthogonal matrix $M$, and $i_\ell, j_\ell \in [d]$ with $\ell \in [2k]$,
\eq{&\E_{M \sim \mathcal O(d)} \left[m(i_1, j_1)\cdots m(i_{2k}, j_{2k}) \right]\\
&=\sum_{p, q \in P_{2k}}
\delta(i_{p(1)}, i_{p(2)}) \cdots \delta(i_{p(2k-1)}, i_{p(2k)})  
\delta(j_{q(1)}, j_{q(2)}) \cdots \delta(j_{q(2k-1)}, j_{q(2k)})
\, \owg(d, (p,q)) \ . 
}
When the number of $m(\cdot, \cdot)$'s is odd, the integral is trivially zero. 

\item 
For a complex normal Gaussian vector $V \in \mathbb C^d$, and $i_\ell, i^\prime_\ell \in [d]$ with $\ell \in [k]$,
\eq{
\E_{V \sim \mathcal G_{\mathbb C}(d)} \left[v(i_1) \cdots v(i_k) \cdot
\overline{v(i^\prime_1) \cdots v(i^\prime_k) } \right]
=\sum_{\alpha \in S_k} 
\delta(i_1, i^\prime_{\alpha (1)}) \cdots \delta(i_k, i^\prime_{\alpha(k)}) 
 \ .
}
When the numbers of $v(\cdot)$'s and $\overline{v(\cdot)}$'s are different, the integral is trivially zero. 

\item 
For a real normal Gaussian vector $V \in \mathbb R^d$, and $i_\ell \in [d]$ with $\ell \in [2k]$,
\eq{
\E_{V \sim \mathcal G_{\mathbb R}(d)} \left[v(i_1)\cdots v(i_{2k}) \right]
=\sum_{p\in P_{2k}}
\delta(i_{p(1)}, i_{p(2)}) \cdots \delta(i_{p(2k-1)}, i_{p(2k)})  
\ .
}
When the number of $v(\cdot)$'s is odd, the integral is trivially zero. 
\end{enumerate} 
\end{theorem}

\begin{corollary}\label{corollary:calculus}
The third and fourth statements of Theorem \ref{theorem:calculus} can be extended to complex and real noraml Gaussian matrices. 
\begin{enumerate}[label={(\arabic*$\,^\prime$)}, start=3]
\item 
For a $d_1 \times d_2$ complex normal Gaussian matrix $M \sim \mathcal G_{\mathbb C}(d_1, d_2)$, and $i_\ell, i^\prime_\ell \in [d_1]$ and $ j_\ell, j^\prime_\ell \in [d_2]$ with $\ell \in [k]$,
\eq{
&\E_{M \sim \mathcal G_{\mathbb C}(d_1d_2)} \left[m(i_1, j_1)\cdots m(i_k, j_k) \cdot
\overline{m(i^\prime_1, j^\prime_1)\cdots m(i^\prime_k, j^\prime_k)} \right]\\
&=\sum_{\alpha \in S_k} 
\delta(i_1, i^\prime_{\alpha (1)}) \cdots \delta(i_k, i^\prime_{\alpha(k)}) 
\cdot \delta(j_1, j^\prime_{\alpha(1)}) \cdots \delta(j_k, j^\prime_{\alpha(k)}) \ .
}
When the numbers of $m(\cdot, \cdot)$'s and $\overline{m(\cdot, \cdot)}$'s are different, the integral is trivially zero. 

\item 
For a $d_1 \times d_2$ real normal Gaussian matrix $M \sim \mathcal G_{\mathbb R}(d_1, d_2)$, and $i_\ell \in [d_1]$ and $ j_\ell \in [d_2]$ with $\ell \in [2k]$,
\eq{
&\E_{M \sim \mathcal G_{\mathbb R}(d_1d_2)} \left[m(i_1, j_1)\cdots m(i_{2k}, j_{2k}) \right]\\
&=\sum_{p \in P_{2k}}
\delta(i_{p(1)}, i_{p(2)}) \cdots \delta(i_{p(2k-1)}, i_{p(2k)})  
\cdot \delta(j_{p(1)}, j_{p(2)}) \cdots \delta(j_{p(2k-1)}, j_{p(2k)})
 \ . 
}
When the number of $m(\cdot, \cdot)$'s is odd, the integral is trivially zero. 
\end{enumerate}
\end{corollary}
Notice that removing Weingarten functions and imposing $\alpha = \beta$ or $p=q$ will make the first two formulae in Theorem \ref{theorem:calculus} into the two in Corollary \ref{corollary:calculus}. Such a correspondence will be exemplified in Section \ref{section:tensornetworks}.

\section{Random matrices in tensor networks}\label{section:tensornetworks}
In this section, we discuss how the integral formulae in Section \ref{section:calculus} can be naturally interpreted in tensor network diagrams, where nodes are tensors and edges represent contractions when both ends are connected to tensors, otherwise they are called dangling edges; see \cite{biamonte2017tensor}. Especially edges connected to no tensors can be regarded as the identities, un-normalized Bell states or their dual operators in the conventional viewpoint of matrix calculus.  

Take a tensor network $T$ having only nodes, i.e. tensors $T^{(1)}, \ldots, T^{(K)}$ whose multi-index sets are individually denoted by $I_1, \ldots, I_K$. Then, $T$ is the tensor product of those tensors, and for $k \in [K]$ pick $i_k \in I_k$ then the $(i_1, \ldots, i_K)$-element is just the product of the corresponding elements;
\eq{
&T = \bigotimes_{k =1}^K T^{(k)} , \quad\quad 
t(i_1, \ldots, i_K) = \prod_{k=1}^K t^{(k)}(i_k) \quad \text{ and } \\
&T = \sum_{\substack{(i_1, \ldots, i_K) \\ \in I_1 \times \cdots \times I_K}}  
t(i_1, \ldots, i_K) \, 
|i_1 \rangle \otimes \cdots \otimes |i_K \rangle \ .
}
Here, without loss of generality, we regard $T$ as a vector using the \emph{ket} notation. 
Next, suppose each multi-index is written as $i_k = (i_{k, 1}, \ldots, i_{k, L(k)})\in I_{k, 1} \times \cdots \times I_{k, L(k)}$ where $L(k)$ can be one. Now, let us introduce an edge to this diagram. For example, if the spaces corresponding to the index sets $I_{k, \ell}$ and $I_{k^\prime, \ell^\prime}$ ($k < k^\prime$) are contracting, where $L(k) = L(k^\prime)$, then we multiply the following dual operator from the left:
\begin{align*}
E = \sum_{i = 1}^{L(k)} \langle i | \otimes \langle i | \otimes I  \ ,
\end{align*}
where the two operators $\langle i|$'s, in the \emph{bra} notation, act on those spaces and the identity the rest. 
That is, the resulting tensor network is:
\eq{
&ET = \\
&\sum_{\substack{(i_1, \ldots, i_K) \\ \in I_1 \times \cdots \times I_K}}  
 \, \delta (i_{k,l}, i_{k^\prime, \ell^\prime}) \cdot 
t(i_1, \ldots, i_K) \, |i_1 \rangle \otimes \cdots \otimes 
|i_k\{\ell\} \rangle \otimes \cdots \otimes |i_{k^\prime}
\{\ell^\prime\} \rangle  \otimes \cdots \otimes  
|i_{K}\rangle  \ .
}
Here, notice that introducing new edges are equivalent to deleting the relevant canonical basis vectors and multiplying delta functions. The same is true with Theorem \ref{theorem:calculus}, where dangling edges which resulted from deleting random tensors will be reconnected, yielding contractions, where newly exposed spaces are to be contracted again for the diagram to balance out.   

Let us apply the above calculus to elementary examples. Consider $MAM^*$ where $A$ is a $d \times d$ deterministic matrix and $M$ is a $d \times d$ random matrix. When $M$ is real, the adjoint operation is just the transpose operation. Now, one can apply Theorem \ref{theorem:calculus} and Corollary \ref{corollary:calculus} to this example with $k=1$, numbering both of $M$ and $M^*$ as $1$ for the complex cases and $M$ and $M^*$ as $1$ and $2$ respectively for the real cases. 
\begin{enumerate}
\item The unitary matrix case: let $\alpha = \beta = \mathrm{id} = (1) \in \mathcal S_1$,
\eq{
\E_{M \sim \mathcal U(d)} \left[m(i_1, j_1)\cdot
\overline{m(i^\prime_1, j^\prime_1)} \right]
&=
\delta(i_1, i^\prime_{\alpha (1)})
\cdot
\delta(j_1, j^\prime_{\beta(1)}) 
\, \uwg(d, (\alpha, \beta)) \quad \text{ or}
\\
\E_{M \sim \mathcal U(d)} \left[m(i, j)\cdot
\overline{m(i^\prime, j^\prime)} \right] 
&= \delta(i, i^\prime) \cdot
\delta(j, j^\prime) \times \frac{1}{n} \ .
}

\item The orthogonal matrix case: let $p = q = \{\{1,2\}\} \in \mathcal P_2$, and
\eq{
\E_{M \sim \mathcal O(d)} \left[m(i_1, j_1)\cdot m(i_2, j_2) \right] 
&=
\delta(i_{p(1)}, i_{p(2)}) \cdot \delta(j_{q(1)}, j_{q(2)}) 
\, \owg(d, (p,q)) \quad \text{ or} \\
\E_{M \sim \mathcal O(d)} \left[m(i, j)\cdot m(i^\prime, j^\prime)\right] 
&=
\delta(i, i^\prime) \cdot \delta(j, j^\prime) 
\times \frac{1}{n} \ .
}
Here, we replaced the set of indices: $i_1, i_2, j_1, j_2$ by $i, i^\prime, j, j^\prime$, respectively.

\item The complex normal Gaussian case can be deduced trivially from the first statement:
\eq{
\E_{M \sim \mathcal G_{\mathbb C}(d,d)}  \left[m(i, j)\cdot
\overline{m(i^\prime, j^\prime)} \right] 
= \delta(i, i^\prime) \cdot \delta(j, j^\prime)
}

\item The real normal Gaussian case can be deduced trivially from the second statement:
\eq{
 \E_{M \sim \mathcal G_{\mathbb R}(d,d)} \left[m(i, j)\cdot m(i^\prime, j^\prime)\right] 
=\delta(i, i^\prime) \cdot \delta(j, j^\prime)
}
Here, we replaced the set of indices: $i_1, i_2$ by $i, i^\prime$, respectively.
\end{enumerate}

\begin{figure}
\centering
\includegraphics[width=\linewidth]{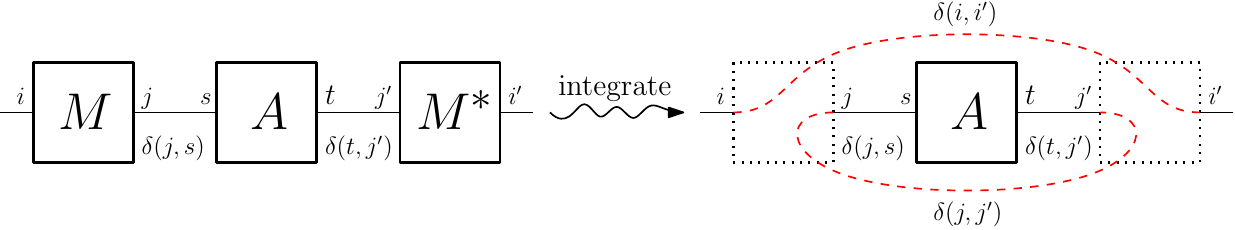}
\caption{
Integrating the diagram, the random matrices $M$ and $M^*$ vanish
and the new dashed red edges appear, which come from delta functions of the integral formulae in Theorem \ref{theorem:calculus} and reconnect newly created dangling edges.} 
\label{fig:trace}
\end{figure}

Thus, the four cases all share the same underlying combinatorics and the integrals are identical apart from the normalization constant. For example, in the complex normal Gaussian case, the $(i, i^\prime)$-element can be calculated as follows. 
\eq{\label{eq:trace}
\left(\E_{M \sim \mathcal G_{\mathbb C}(d,d)}[MAM^*]\right)_{i, i^\prime}
&= \sum_{j,s,t,j^\prime} a(s, t) \cdot  \delta(j,s) \cdot \delta(t, j^\prime)
\E_{M \sim \mathcal G_{\mathbb C}(d,d)}  \left[m(i, j)\cdot
\overline{m(i^\prime, j^\prime)} \right] \\
&=  \sum_{j,s,t,j^\prime} a(s, t) \cdot  \delta(j,s) \cdot \delta(t, j^\prime) \cdot \delta(i, i^\prime) \cdot \delta(j, j^\prime) \ ,
}
which amounts to 
\eq{
\E_{M \sim \mathcal G_{\mathbb C}(d, d)}[MAM^*] = \trace[A]I \ .
}
In the above calculations, matrix multiplications are realized by the delta functions, which explains the delta functions appearing in Figure \ref{fig:trace}. This integration process can be described by ``deleting random tensors and reconnecting the newly created dangling edges.''

\begin{figure}[h]
  \centering
  \begin{subfigure}[b]{\textwidth}
    \centering
    \includegraphics[width=1.0\textwidth]{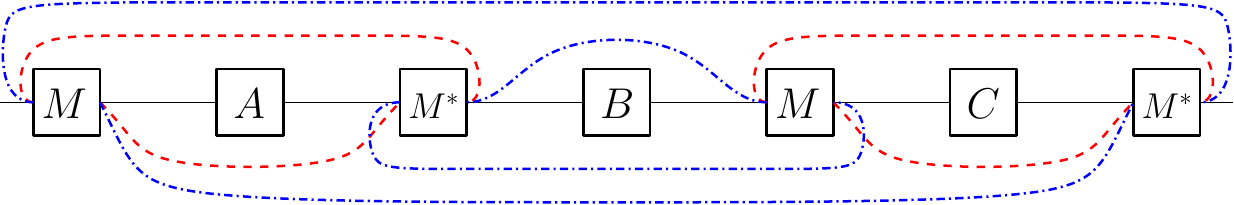}
    \caption{Complex cases.}
    \label{fig:complex}
  \end{subfigure}
  
  \vspace{5mm} 
  
  \begin{subfigure}[b]{\textwidth}
    \centering
    \includegraphics[width=1.0\textwidth]{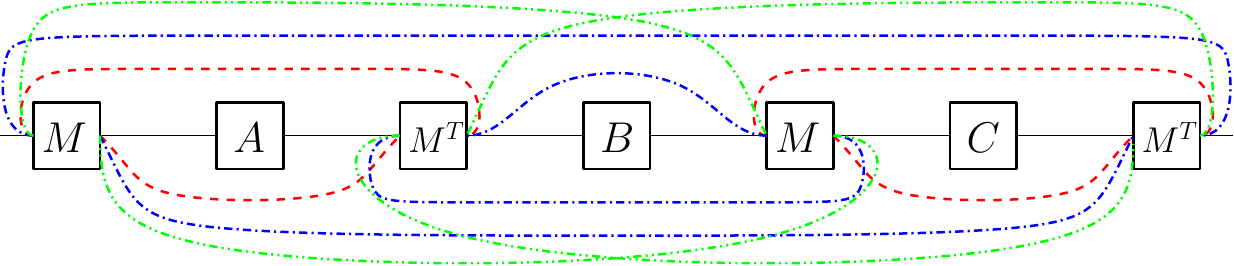}
    \caption{Real cases.}
    \label{fig:real}
  \end{subfigure}
  
  \caption{Reconnecting the diagram. First, there are two different cases - complex and real, as shown in \ref{fig:complex} and \ref{fig:real}.
  In each diagram, the top side corresponds to the output space of $M$ and the bottom the input space. The dashed red edges indicate the reconnections caused by $(1)(2)$ and $\{\{1,2\},\{3,4\}\}$. Similarly the dash-dotted blue edges correspond to $(1,2)$ and $\{\{1,4\},\{2,3\}\}$ and the dash-dot-dotted green edges to $\{\{1,3\},\{2,4\}\}$. Here, the copies of $M$ (and $M^*$) are numbered from left to right. 
  } 
  \label{fig:reconnection}
\end{figure}

Now, in order to see the fundamental differences among various random tensors, let us calculate 
\begin{align}
\E[MAM^*BMCM^*] \ ,
\end{align}
where $A, B, C$ are deterministic matrices and $M$ is a random matrix. Table \ref{table:example} shows all possible configurations. Similarly as before, $M$s and $M^*$s are numbered from left to right, and $\alpha, \beta \in S_{2}$ and $p, q \in P_4$. Young diagrams corresponding to pairs $(\alpha, \beta)$ or $(p, q)$ define the relevant Weingarten functions as explained in Appendix \ref{section:weingarten}. The resulting tensor network diagrams, after reconnection, are listed on Table \ref{table:example}, and Figure \ref{fig:reconnection} explains how they were obtained. Let us consider two configurations as examples. First, the unitary matrix case with $\alpha = (1,2), \beta = (1)(2)$ is shown in the third row of Table \ref{table:example} and represented by the blue dash-dotted edges on the top and the red dashed edges on the bottom in Figure \ref{fig:complex}. Going along these edges yields $\trace [A] \trace[C] \trace[B] \, I$. Following Appendix \ref{section:weingarten}, $\alpha^{-1} \beta = (1,2)$ gives the cycle structure $(2)$, which is the corresponding Young diagram, so that $\uwg(d, (2))$ in \eqref{eq:small-size-unitary-example} must be multiplied. 
Thus, this configuration makes the following contribution to the integral:
\eq{
\frac{-1}{d(d^2-1)} \, \trace [A] \trace[C] \trace[B] \, I \ .
}
Next, the real normal Gaussian case with $p=q=\{\{1,3\}, \{2,4\}\}$ is shown in the last row of Table \ref{table:example} and represented by the green dash-dot-dotted edges on the top and bottom in Figure \ref{fig:complex}. Remember that for real normal Gaussian tensors, $p=q$ always holds ($\alpha=\beta$ for the complex normal Gaussian tensors), and such cases are shaded on the table. As before, following these edges yields 
\eq{
\trace[AC^T]\, B \ ,
}
which is the contribution of this configuration. If it was orthogonal, one would need to specify the Weingarten function additionally as follows. Using the formula in Appendix \ref{section:weingarten}, let $\tilde p = \tilde q = (1,3)(2,4)$ so that $\tilde p^{-1}q = (1)(2)(3)(4)$ gives the cyclic structure $(1,1)$ - the corresponding Young diagram. Hence, multiplying the above diagram by $\owg(d, (1,1))$ in \eqref{eq:small-size-orthogonal-example},
 the contribution to the integral would be
\eq{
\frac{-1}{d(d-1)(d+2)} \, \trace [AC^T] \, B \ .
}

\begin{table} [h] 
  \centering
\renewcommand{\arraystretch}{1.2}
  \begin{tabular}{|c|c|c||c|c ||c|c|}
    \hline
    &$\alpha$ & $\beta$ & $p$ & $q$ & \small Young Diagram & \small Tensor Network Diagram \\
    \hline 
    \multirow{4}{*}{\rotatebox{90}{\hspace{-1.5mm}{\tiny Complex \& Real}}}
    & \cellcolor{gray!20}$(1)(2)$ &\cellcolor{gray!20} $(1)(2)$ &\cellcolor{gray!20}$\{\{1,2\}, \{3,4\}\}$ & \cellcolor{gray!20}$\{\{1,2\}, \{3,4\}\}$  & (1,1) &  $\trace [A] \trace[C] \, B$ \\
    \cline{2-7}
    &$(1)(2)$ & $(1,2)$ & $\{\{1,2\}, \{3,4\}\}$ & $\{\{1,4\}, \{2,3\}\}$ & (2) & $\trace [AC] \, B$ \\
    \cline{2-7}
    &$(1,2)$ & $(1)(2)$ & $\{\{1,4\}, \{2,3\}\}$ & $\{\{1,2\}, \{3,4\}\}$ & (2) & $\trace [A] \trace[B] \trace[C] \, I$ \\
    \cline{2-7}
    &\cellcolor{gray!20}$(1,2)$ & \cellcolor{gray!20}$(1,2)$ & \cellcolor{gray!20}$\{\{1,4\}, \{2,3\}\}$ & \cellcolor{gray!20}$\{\{1,4\}, \{2,3\}\}$ & (1,1) &  $\trace [AC] \trace[B] \, I$ \\
    \cline{1-7} \multicolumn{7}{|c|}{} \\[-4.5mm] \cline{1-7}
    \multirow{5}{*}{\rotatebox{90}{{\tiny Real only}}}
    &N/A & N/A & $\{\{1,2\}, \{3,4\}\}$ & $\{\{1,3\}, \{2,4\}\}$  & (2) &  $\trace [AC^T] \, B$ \\
    \cline{2-7}
    &N/A & N/A & $\{\{1,4\}, \{2,3\}\}$ & $\{\{1,3\}, \{2,4\}\}$  & (2) &  $\trace [AC^T] \trace[B] \, I$ \\
    \cline{2-7}
    &N/A & N/A & $\{\{1,3\}, \{2,4\}\}$ & $\{\{1,2\}, \{3,4\}\}$  & (2) &  $\trace [A] \trace[C] \,  B^T$ \\
    \cline{2-7}
    &N/A & N/A & $\{\{1,3\}, \{2,4\}\}$ & $\{\{1,4\}, \{2,3\}\}$  & (2) &  $\trace [AC] \, B^T$ \\
    \cline{2-7}
    &N/A & N/A & \cellcolor{gray!20}$\{\{1,3\}, \{2,4\}\}$ & \cellcolor{gray!20}$\{\{1,3\}, \{2,4\}\}$  & (1,1) &  $\footnotesize \trace [AC^T] \,  B^T$ \\
    \hline
  \end{tabular}  
\caption{Combinatorics and tensor network diagrams.}
\label{table:example}
\end{table}

The above examples bear only simple calculations in the sense that the output and the input spaces of $M$ are independent of each other while reconnecting the diagram. Indeed, $p$ has three possibilities: $\{\{1,2\}, \{3,4\}\}$, $\{\{1,4\}, \{2,3\}\}$ and $\{\{1,3\}, \{2,4\}\}$, but they give respectively $B$, $\trace [B] \, I$ and $B^T$. Similarly, the three choices for $q$ yield $\trace[A]\trace[C]$, $\trace[AC]$ and $\trace[AC^T]$. However, in general the edges will be intertwined. Moreover, the resulting diagram might contain loops with no tensor, each of which will contribute the factor of $\trace I$. 

Finally we briefly discuss how \texttt{PyRTNI2} treats random tensors. In the above example, it was crucial to treat $M$s and $M^*$s together. More generally, for a random tensor $M$, the four variants $M$, $\bar M$, $M^T$ and $M^*$ must be handled together; the last two apply only to matrices. For this reason, \texttt{PyRTNI2} creates a tensor based on a name. Once a tensor was created, it can be cloned and one of the three operations: transpose, complex conjugate, adjoint can be applied if possible and necessary. In general, tensors are associated with multiple spaces, which are numbered internally. In \texttt{PyRTNI2}, tensors are more fundamental than matrices, so \texttt{PyRTNI2} provides a wrapper to treat a tensor as a matrix, where the spaces of the tensor will belong to either the output or the input, and will be numbered individually. This numbering system is used to create edges in tensor network diagrams. Other deterministic matrices and tensors can be created in the same way. In fact, one sets a tensor or a matrix to be random just before the symbolic integration, one after another, where
different kinds of random tensors can be placed in the same diagram.
See tutorial files on GitHub for the detail.

\section{Further calculus}\label{section:further}
\subsection{Tensored completely positive maps}\label{section:cp}
In \cite{RTNI}, the following example was treated: using the bra-ket notation, 
\eq{\label{eq:tensored_channels}
\E_{U \sim \mathcal U (k\ell)} \left[\langle \omega_k |\trace_{\mathbb C^\ell}[(U \otimes \bar U)\, (|\omega_m \rangle \langle \omega_m| \otimes |0 \rangle \langle 0|)\,(U^* \otimes \bar U)] |\omega_k \rangle \right] \ ,
}
where
$ 
| \omega_d \rangle = \sum_{i=0}^{d-1} |i\rangle 
$ is an unnormalized Bell state
and $\{|i \rangle \}_{i=1}$ is the canonical basis for $\mathbb C^d$. There is no vector equivalent to $|0\rangle$ in \cite{RTNI} because they all will get contracted to be $\langle 0|0 \rangle=1$ anyway. Note, however, that $U (I_m \otimes |0\rangle)$ is an isometry mapping from $\mathbb C^{k\ell/m}$ to $\mathbb C^{k\ell}$. In this sense, $m$ morally must divide $k\ell$, but one can drop the above $|0\rangle$ and ignore the condition. As in \cite{RTNI} for the detail, the integral in \eqref{eq:tensored_channels} amounts to
\eq{
&\frac{1}{(k\ell)^2-1} \left[
\trace [I_k] \, (\trace [I_\ell])^2 \, \trace [I_m]
+ (\trace [I_k])^2 \, \trace [I_\ell] \, (\trace [I_m])^2
\right]\\
&\qquad - \frac{1}{(k\ell)^3-(k\ell)^2}\left[ 
\trace [I_k] \, (\trace [I_\ell])^2 \, (\trace [I_m])^2
+ (\trace [I_k])^2 \, \trace [I_\ell] \, \trace [I_m]
\right]\\
&= \frac{k \ell^2 m + k^2\ell m^2}{(k\ell)^2-1} 
- \frac{k \ell^2 m^2 + k^2 \ell m}{(k\ell)^3-(k\ell)^2}
}
Notice that $k\ell$ the output dimension of $U$ was used to write the Weingarten functions, and \texttt{PyRTNI2} will do the same by default, although one can switch the reference dimension to the input space. 

\begin{figure}[h]
\centering
\includegraphics[width=\linewidth]{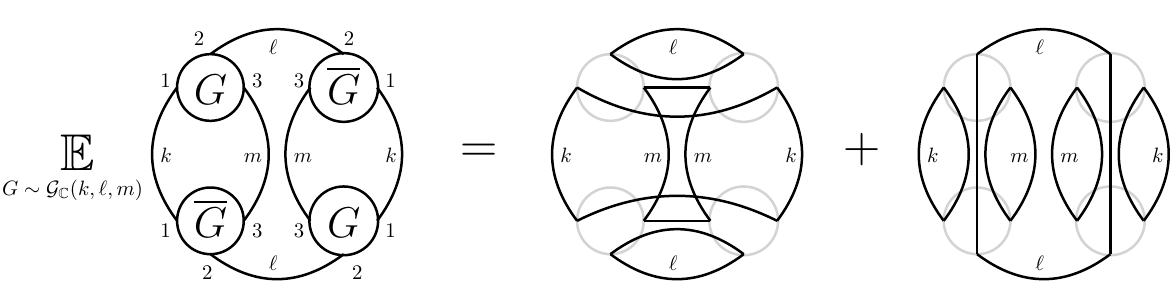}
\caption{
The LHS is the diagrammatic representation of \eqref{eq:tensored_cps}. There are no transpose or adjoint operations, and only the numbers $1,2,3$ are assigned respectively to the  spaces $\mathbb C^k, \mathbb C^l, \mathbb C^m$.
Reconnecting the left diagram, one gets the integrals as the sum of the two diagrams on the RHS, i.e. $k \ell^2 m + k^2 \ell m^2$.
}
\label{fig:cp}
\end{figure}

Now, let us replace the unitary matrix $U$ by a complex normal Gaussian matrix $G$. To present the new formula as an mathematical expression, we regard $G$ as a $k\ell \times m$ matrix in the following:
\eq{\label{eq:tensored_cps}
\E \left[\langle \omega_k |\trace_{\mathbb C^\ell}[(G \otimes \bar G) |\omega_m \rangle \langle \omega_m|(G^* \otimes  G^T
)] |\omega_k \rangle \right] \ .
}
However, the notions of the input space and the output space are not essential for real and complex normal Gaussian tensors. From the tensor diagrammatic viewpoint, we just can assign the id numbers $1,2,3$ respectively to $\mathbb C^k, \mathbb C^l, \mathbb C^m$, as in Figure \ref{fig:cp}, where there are no transpose or adjoint operations. To integrate the tensor network, assign the numbers $1,2$ to $G$ and $\bar G$ individually, from the top to the bottom, and reconnect the diagram, which are comparable to the top two shaded rows of Table \ref{table:example}. That is
\eq{
\trace[I_k] \, (\trace [I_\ell])^2 \, \trace[I_m]
+ (\trace[I_k])^2 \, \trace [I_\ell] \, (\trace[I_m])^2
= k \ell^2 m + k^2 \ell m^2 \ .
}

\subsection{Partially transposed Wishart matrices}\label{section:wishart}

Wishart matrices have been investigated for many years since it was formulated in \cite{wishart} and play an important role in multivariate statistics. 
A Wishart matrix is made of two Gaussian matrices:
\eq{
W = GG^* \ .
}
As before, when $G$ is real, the adjoint operation is understood as the transpose operation. Although, the moments have been already calculated under various conditions, see for example \cite{real_wishart_moment, complex_wishart_moment}, one can reproduce them using \texttt{PyRTNI2} for elementary cases. 

\begin{figure}[h]
\centering
\includegraphics[width=0.9\linewidth]{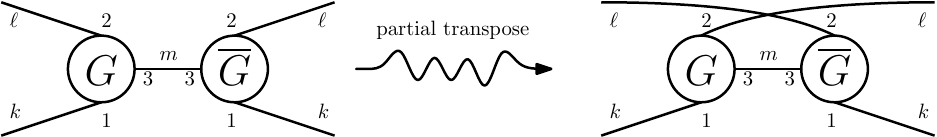}
\caption{
A partially transposed complex Wishart matrix. The left diagram is a $k\ell \times k\ell$ Wishart matrix with the contracted space being $m$-dimensional.
On the right diagram is the left diagram is partially transposed only on the $\ell$-dimensional space. 
As before, the three spaces of $k, \ell, m$ dimensional spaces are respectively numbered as $1,2,3$ for the purpose of reconnection. 
A real partially transposed Wishart matrix can be constructed in a similar way, i.e. literally dropping the complex conjugate notation. 
}
\label{fig:wishart_pt}
\end{figure}

Unlike Wishart matrices, 
partially transposed complex Wishart matrices were investigated rather recently \cite{Aubrun2012,
BanicaNechita2013,
FS2013,
mingo2019freeness,
mingo2022partial}, inspired by the entanglement test in quantum information theory. Partially transposed complex Wishart matrices are diagrammatically constructed in Figure \ref{fig:wishart_pt}, where the size of the Wishart matrix is $k\ell \times k\ell$, the contracting space is $m$-dimensional and the partial transpose was applied on the $\ell$-dimensional space.
Partially transposed real Wishart matrices can be similarly composed. 
Now, we present the low moments calculated by \texttt{PyRTNI2}. 
\begin{itemize}
\item The moments of partially transposed real Wishart matrices:
\begin{enumerate}
\item the first moment:
\begin{dmath*}
k \ell m
\end{dmath*}
\item the second moment:
\begin{dmath*}
k^{2} \ell^{2} m + k \ell m^{2} + k \ell m
\end{dmath*}
\item the third moment:
\begin{dmath*}
k^{3} \ell m + 3 k^{2} \ell^{2} m^{2} + 3 k^{2} \ell m + k \ell^{3} m + 3 k \ell^{2} m + k \ell m^{3} + 3 k \ell m^{2}
\end{dmath*}
\item the fourth moment:
\begin{dmath*}
 k^{4} \ell^{2} m + 2 k^{3} \ell^{3} m^{2} + 2 k^{3} \ell^{2} m + 4 k^{3} \ell m^{2} + 4 k^{3} \ell m + k^{2} \ell^{4} m + 2 k^{2} \ell^{3} m + 6 k^{2} \ell^{2} m^{3} + 5 k^{2} \ell^{2} m^{2} + 10 k^{2} \ell^{2} m + 12 k^{2} \ell m^{2} + 8 k^{2} \ell m + 4 k \ell^{3} m^{2} + 4 k \ell^{3} m + 12 k \ell^{2} m^{2} + 8 k \ell^{2} m + k \ell m^{4} + 6 k \ell m^{3} + 5 k \ell m^{2} + 8 k \ell m 
\end{dmath*}
\item the fifth moment:
\begin{dmath*}
k^{5} \ell m + 10 k^{4} \ell^{2} m^{2} + 10 k^{4} \ell m + 10 k^{3} \ell^{3} m^{3} + 10 k^{3} \ell^{3} m + 25 k^{3} \ell^{2} m^{2} + 30 k^{3} \ell^{2} m + 10 k^{3} \ell m^{3} + 30 k^{3} \ell m^{2} + 25 k^{3} \ell m + 10 k^{2} \ell^{4} m^{2} + 25 k^{2} \ell^{3} m^{2} + 30 k^{2} \ell^{3} m + 10 k^{2} \ell^{2} m^{4} + 25 k^{2} \ell^{2} m^{3} + 70 k^{2} \ell^{2} m^{2} + 70 k^{2} \ell^{2} m + 30 k^{2} \ell m^{3} + 70 k^{2} \ell m^{2} + 60 k^{2} \ell m + k \ell^{5} m + 10 k \ell^{4} m + 10 k \ell^{3} m^{3} + 30 k \ell^{3} m^{2} + 25 k \ell^{3} m + 30 k \ell^{2} m^{3} + 70 k \ell^{2} m^{2} + 60 k \ell^{2} m + k \ell m^{5} + 10 k \ell m^{4} + 25 k \ell m^{3} + 60 k \ell m^{2} + 52 k \ell m
\end{dmath*}
\end{enumerate}
\item The moments of partially transposed complex Wishart matrices:
\begin{enumerate}
\item the first moment:
\begin{dmath*}
k \ell m
\end{dmath*}
\item the second moment:
\begin{dmath*}
k^{2} \ell^{2} m + k \ell m^{2}
\end{dmath*}
\item the third moment:
\begin{dmath*}
k^{3} \ell m + 3 k^{2} \ell^{2} m^{2} + k \ell^{3} m + k \ell m^{3}
\end{dmath*}
\item the fourth moment:
\begin{dmath*}
k^{4} \ell^{2} m + 2 k^{3} \ell^{3} m^{2} + 4 k^{3} \ell m^{2} + k^{2} \ell^{4} m + 6 k^{2} \ell^{2} m^{3} + 4 k^{2} \ell^{2} m + 4 k \ell^{3} m^{2} + k \ell m^{4} + k \ell m^{2}
\end{dmath*}
\item the fifth moment:
\begin{dmath*}
k^{5} \ell m + 10 k^{4} \ell^{2} m^{2} + 10 k^{3} \ell^{3} m^{3} + 10 k^{3} \ell^{3} m + 10 k^{3} \ell m^{3} + 5 k^{3} \ell m + 10 k^{2} \ell^{4} m^{2} + 10 k^{2} \ell^{2} m^{4} + 30 k^{2} \ell^{2} m^{2} + k \ell^{5} m + 10 k \ell^{3} m^{3} + 5 k \ell^{3} m + k \ell m^{5} + 5 k \ell m^{3} + 2 k \ell m
\end{dmath*}
\end{enumerate}
\end{itemize}

\section*{Acknowledgment}
MF acknowledges JSPS KAKENHI Grant Number JP20K11667. This work was supported by Japan-France Integrated action Program
(SAKURA), Grant number JPJSBP120203202. 
MF thanks Sho Matsumoto for answering questions on orthogonal Weingarten functions, and Benoit Collins and Ion Nechita for fruitful discussions.

\appendix

\section{Generating Weingarten functions}\label{section:weingarten}
In this section, we explain the complex and real Weingarten functions, whose values depend on the matrix sizes and Young diagrams, although pairs of permutations or pairs of pair partitions replaced Young diagrams to avoid complication in the main body. 

The unitary Weingarten functions are defined to Young diagrams: abusing the notation, 
\begin{align}
\uwg(d, \mu) &= \uwg(d, (\alpha, \beta)) \quad \text{ for } \alpha, \beta \in S_k \ ,
\end{align}
where the map $(\alpha, \beta) \mapsto \mu$ is defined by the lengths, in the non-increasing order, of loops of $\alpha^{-1}\beta$,
say, $\mu_1 \geq \mu_2 \geq \cdots$, giving a Young diagram. Then \cite{CollinsSniady2006}, 
\begin{align}\label{eq:weingarten-unitary}
\uwg(d, \mu) = \frac{1}{(k !)^2} \sum_{\substack{\lambda \vdash k \\ \ell(\lambda) \leq d}} \frac{\chi^\lambda(e)^2}{s_{\lambda, d} (1)} \, \chi^{\lambda}(\mu) \ .
\end{align}
Here, 
\begin{itemize}
\item The sum runs over all Young diagrams $\lambda$'s of $k$ elements whose lengths $\ell(\lambda)$ are not more than $d$.
\item $\chi^{\lambda}$ is the irreducible character of $\mathcal S_k$, 
and $e $ is the identity element of $\mathcal S_k$. 
\item $s_\lambda(1)$ is the Schur polynomial with respect to $\lambda$:
\begin{align}
s_\lambda = \prod_{1 \leq i < j \leq k} \frac{\lambda_i - \lambda_j + j -i}{j-i} \ .
\end{align}
\end{itemize}

Let us write down unitary Weingarten functions for $k = 1, 2$:
\eq{\label{eq:small-size-unitary-example}
\uwg(d, (1)) = \frac{1}{d}, \quad \uwg(d, (1, 1)) = \frac{1}{d^2-1}, \quad \uwg(d, (2)) = \frac{-1}{d(d^2-1)} \ .
}

The orthogonal Weingarten functions are also defined with respect to individual Young diagrams: again abusing the notation, 
\begin{align}
\owg(d, \mu) &= \owg(d, (p, q)) \quad \text{ for } p, q \in P_{2k} \ .
\end{align}
where the map $(p, q) \mapsto \mu$ is defined as follows. 
Consider the graph $\Gamma(p, q)$ with vertices $[2k]$ and with edges $\{p(2\ell -1), p(2 \ell)\}$ and $\{q(2\ell -1), q(2 \ell)\}$ for $\ell \in [k]$. Then, the graph consists of loops with lengths, for example, $2 \mu_1 \geq 2\mu_2 \geq \cdots$, resulting in the Young diagram $(\mu_1, \mu_2, \ldots)$ of $k$. To see this, place the vertices $[2k]$ on the horizontal straight line, and draw the edges by $p$ over the line and those by $q$ under. This view gives another way of obtaining the Young diagrams, i.e. just regard $p$ and $q$ as transpositions in $S_{2k}$ such that 
\eq{
\tilde p = (p(1), p(2))(p(3), p(4))\cdots \qquad \tilde q = (q(1), q(2))(q(3), q(4))\cdots \ .
}
Then, the loops of $\tilde p^{-1} \tilde q$ have the following lengths: $\mu_1, \mu_1, \mu_2, \mu_2,\ldots$, where the vertices of each loop of $\Gamma(p, q)$ are exactly the same as those of certain two loops of $\tilde p^{-1} \tilde q$. Moreover, notice that the edges of $\Gamma(p, q)$ can be replaced by $\{2\ell -1, 2 \ell\}$ and $\{p^{-1}q(2\ell -1), p^{-1}q(2 \ell)\}$ permuting all vertices by $p^{-1}$.
Finally, 
orthogonal Weingarten functions can be written as follows \cite{CollinsMatsumoto2009}.
\begin{align}\label{eq:weingarten-orthogonal}
\owg(d, \mu) = \frac{2^k k!}{(2k)!} \sum_{\substack{\lambda \vdash k \\ \ell(\lambda) \leq d}}  
\frac{\chi^{2\lambda} (e)^2 }{Z_\lambda(1)} \omega^\lambda (\mu) \ .
\end{align} 
Here,
\begin{itemize}
\item The sum is made over Young diagrams $\lambda$'s of $k$ whose lengths $\ell(\lambda)$ are not more than $d$.
\item $\chi^{2\lambda}$ is the irreducible character of $\mathcal S_{2k}$ where $2 \lambda$ is defined as $(\lambda_i)_i \mapsto (2\lambda_i)_i$. Also, $e$ is the identity element of $\mathcal S_{2k}$.
\item $Z_\lambda (1)$ is the zonal polynomial:
\begin{align}
Z_\lambda(1) = \prod_{(i,j) \in \lambda} (d + 2j - i -1) \ .
\end{align}
\end{itemize}

Let us write down orthogonal Weingarten functions for $k=1,2$:
\eq{\label{eq:small-size-orthogonal-example}
\owg(d, (1)) = \frac{1}{d}, \quad \owg(d, (1, 1)) = \frac{d+1}{d(d-1)(d+2)}, \quad \owg(d, (2)) = \frac{-1}{d(d-1)(d+2)} \ .
}

\begin{remark}\label{remark:low-dim}
Notice that each of \eqref{eq:weingarten-unitary} and \eqref{eq:weingarten-orthogonal} gives different formulae for individual pairs $d<k$ because Young diagrams $\lambda$'s whose length are more than $d$ do not appear in the sum. Otherwise, 
formally setting $d=1$ in \eqref{eq:small-size-unitary-example} and 
\eqref{eq:small-size-orthogonal-example}, $\uwg(1, (1,1))$, $\uwg(1, (2))$, $\owg(1, (1,1))$ and $\owg(1, (2))$ would explode. 
\end{remark}

\bibliographystyle{alpha}
\bibliography{reference}{}

\end{document}